\documentclass[10pt,aps,prd,reprint,nofootinbib]{revtex4-1}

\usepackage[utf8]{inputenc}
\usepackage{graphicx}
\usepackage{amsmath}
\usepackage[colorlinks,pdfusetitle]{hyperref}
\hypersetup{colorlinks=true,allcolors=[rgb]{1,0.56,0}}

\newcommand{\orcid}[1]{\href{https://orcid.org/#1}{#1}}

\newcommand{\e}[1]{\times10^{#1}}

\newcommand{\vev}[1]{\langle {#1} \rangle}
\newcommand{\lsim}{\lesssim}
\newcommand{\gsim}{\gtrsim}
\newcommand{\eq}[1]{Eq.~(\ref{#1})}
\newcommand{\ord}[1]{\mathcal{O}{(#1)}}
\newcommand{\beq}{\begin{equation}}
\newcommand{\eeq}{\end{equation}}

\newcommand{\phie}{\phi_\oplus}
\newcommand{\trh}{T_{\rm rh}}
\newcommand{\lphi}{\lambda_\phi}

\begin{document}

\title{Sterile Neutrino Shape-shifting Caused by Dark Matter}

\author{Hooman Davoudiasl}
\email{hooman@bnl.gov}
\thanks{\orcid{0000-0003-3484-911X}}

\author{Peter B.~Denton}
\email{pdenton@bnl.gov}
\thanks{\orcid{0000-0002-5209-872X}}

\affiliation{High Energy Theory Group, Physics Department, Brookhaven National Laboratory, Upton, NY 11973, USA}

\begin{abstract}
Light sterile neutrinos with a mass of $\sim 1$ eV continue to be interesting due to multiple hints from terrestrial experiments.
This simple hypothesis suffers from strong astrophysical constraints, in particular from the early universe as well as solar neutrinos.
We develop a cosmologically viable proposal consistent with the terrestrial hints, as well as solar constraints, by sourcing the sterile neutrino's mass from ordinary matter via an ultralight scalar $\phi$ which can also be the dark matter. In this scenario, the experimentally implied $\sim 1$ eV sterile neutrino mass is a local value and changes throughout spacetime.
\end{abstract}

\maketitle

\section{Introduction}
The three-flavor neutrino picture is coming into view with oscillation data from many experiments.
Some key tensions remain, many of which can be addressed by the presence of a new light sterile neutrino with a mass $\sim1$ eV and relatively large mixing \cite{Diaz:2019fwt,Boser:2019rta,Giunti:2019aiy,Acero:2022wqg}.
This explanation, however, is in tension with a number of data sets, notably cosmological data from the early universe \cite{Hannestad:2010kz,Wong:2011ip,Lattanzi:2017ubx,Hagstotz:2020ukm}, as well as those from terrestrial and solar neutrino experiments \cite{Dentler:2018sju}.

In this article, we propose a solution to some of the constraints on light sterile neutrinos by dynamically changing the mass of the sterile neutrino depending on its environment -- an ultralight depth-dependent scalar non-standard interaction in the sterile sector, see e.g.~Refs.~\cite{Ge:2018uhz,Wise:2018rnb,Denton:2018dqq,Babu:2019iml,Smirnov:2019cae,Venzor:2020ova,Medhi:2021wxj,Medhi:2022qmu,Denton:2022pxt} for other papers with components of this model.
This is accomplished by coupling the sterile neutrino to an ultralight scalar which in turn is coupled to baryons.
This model evades cosmological constraints as well as solar constraints and the scalar can be a viable candidate for dark matter (DM) depending on the nature of the scalar's self-interaction as will be discussed.
We will provide concrete viable details, but many of the details can be modified for similar, or even richer, phenomenology.

Some aspects of our scenario have been discussed before with key differences such as the lack of a DM candidate \cite{Dasgupta:2013zpn,Escudero:2022gez,Ghalsasi:2016pcj}, and the lack of an explanation for the $\sim1$ eV sterile neutrino hints \cite{Gonzalez-Garcia:2006vic,Dev:2022bae}.
In addition, none address the constraint from solar neutrinos.
Other models looking to reconcile $\sim1$ eV sterile neutrinos with cosmology often turn to neutrino self-interactions \cite{Hannestad:2013ana,Dasgupta:2013zpn}, although it is known that this introduces more problems \cite{Song:2018zyl,Chu:2018gxk}; such a scenario could partially improve \cite{Kreisch:2019yzn} the Hubble tension \cite{Planck:2018vyg,Riess:2019cxk,Abdalla:2022yfr}.
Other studies have investigated the connection between ultralight particles and neutrino oscillation experiments \cite{Berlin:2016woy,Brdar:2017kbt,Krnjaic:2017zlz,Davoudiasl:2018hjw,Liao:2018byh,Capozzi:2018bps,Huang:2018cwo,Farzan:2019yvo,Cline:2019seo,Dev:2020kgz,Losada:2021bxx,Huang:2021kam,Chun:2021ief,Huang:2022wmz} or DM \cite{Davoudiasl:2022ubh}.

\section{Light Sterile Neutrinos}
A new sterile neutrino is an economical way to develop rich neutrino oscillation phenomenology and has been a standard benchmark for many interesting theoretical and experimental questions and may well be related to the nature of neutrino mass.
A number of anomalous data sets regularly suggest $m_4\sim1$ eV which typically mixes either with $\nu_e$ or both $\nu_e$ and $\nu_\mu$ at the $\sim1\%$ to $\sim10\%$ level.

In the case where $\nu_4$ primarily mixes with $\nu_e$, the most significant hint comes from gallium data \cite{Giunti:2010zu,SAGE:2009eeu,Kaether:2010ag}.
While the theory involved is somewhat complicated, no explanation within standard physics seems to exist \cite{Kostensalo:2019vmv,Brdar:2023cms,Haxton:2023ppq} and a recent experiment claimed a high significance result at the $>5\sigma$ level at $m_4\gtrsim1$ eV and $\sin^22\theta_{ee}\sim0.4$ \cite{Barinov:2021asz,Giunti:2022xat}.
These parameters are in tension with solar data using theoretical predictions for the solar neutrino flux \cite{Goldhagen:2021kxe}.

Additional weak hints come from short-baseline accelerator data and reactor neutrinos compared to theory or reactor spectral data \cite{Mention:2011rk,T2K:2014xvp,Berryman:2020agd,Denton:2021czb} which may also point towards a similar picture.
For these data sets either other explanations exist or the significance is quite low \cite{DayaBay:2017jkb,RENO:2018pwo,Arguelles:2021meu,MicroBooNE:2022sdp}\footnote{To be specific, one $\nu_e$ disappearance analysis of the MicroBooNE data found weak evidence $2.4\sigma$ of $\nu_e$ disappearance due to a dip in the data that is consistent with the parameters for the gallium data \cite{Denton:2021czb}, other analyses found that this dip is less significant \cite{Arguelles:2021meu,MicroBooNE:2022sdp}.}.
Reactor spectral data also disfavors the large mixing preferred by some anomalies \cite{Berryman:2020agd}.

The additional inclusion of $\nu_\mu$ mixing has the strengths (and weaknesses) of mixing with just $\nu_e$ and is impacted by several new important data sets.
Notably LSND \cite{LSND:2001aii} and MiniBooNE \cite{MiniBooNE:2020pnu} have each seen evidence for $\nu_\mu\to\nu_e$ mixing at $3.8\sigma$ and $4.8\sigma$ respectively that, when interpreted with other constraints, leads to a preferred region of parameter space of $m_4\sim0.7$ eV and $\sin^22\theta_{\mu e}\sim0.01$ \cite{Dentler:2018sju}.
These parameters are also tested elsewhere notably MINOS+ \cite{MINOS:2017cae} and IceCube \cite{IceCube:2020phf} which have derived constraints that put the high significance results from LSND and MiniBooNE into confusing tension.
Simple extensions to the $\sim1$ eV sterile neutrino picture such as multiple sterile neutrinos in this range \cite{Kopp:2013vaa,Diaz:2019fwt} or new interactions \cite{Liao:2016reh,Denton:2018dqq} do not significantly clarify this picture.

In any case, strong constraints from cosmology exist which are largely flavor independent and strongly disfavor all of the interesting parameter spaces discussed above \cite{Hagstotz:2020ukm}.

\section{An Ultralight Boson Model}

We will consider the Lagrangian for the interactions of the ultralight scalar $\phi$ with the sterile neutrino $\nu_s$ and Earth, {\it i.e.}~electrons $e$ and nucleons $n$
\beq
{\cal L} \supset -(m_0+ g_s \phi)\, \bar \nu_s \nu_s - g_e \phi\, \bar e e - g_n \phi\,\bar n n\,,
\label{L}
\eeq
where $m_0$ is a bare mass term.
In Appendix~\ref{sec:uvphysics}, we outline possible underlying ultraviolet (UV) structures and symmetries that could explain our choice of parameters in this section, in general terms.

A long range force acting on electrons and nucleons is constrained to have strength $g_e \lsim 1.4 \times 10^{-25}$ and $g_n \lsim 8.0\times 10^{-25}$ at $2\sigma$, respectively \cite{Fayet:2017pdp,MICROSCOPE:2022doy}.
For simplicity, we will examine the case where $g_e\ll g_n$ and thus electron densities can be ignored, although including $g_e$ as well would not significantly impact our discussion.

The potential for $\phi$ is assumed to be given by 
\beq
V(\phi) = \frac{1}{2} m_\phi^2 \phi^2 + 
\frac{\lphi}{4!} \phi^4\,,
\label{Vphi}
\eeq
where $m_\phi$ is the mass of $\phi$ and $\lphi$ denotes its self-coupling strength.
We will posit that the mass\footnote{We keep $g_sg_n<0$ to ensure that $m_s$ never passes through zero.} implied by laboratory experiments for $\nu_4$, which is mostly composed of $\nu_s$, is sourced coherently by the terrestrial nucleons (baryons) in addition to the bare mass term: $m_s \approx m_0+g_s \phi$.
We will consider $m_\phi$ smaller than the inverse of Earth's radius $1/R_\oplus \approx (6400~\text{km})^{-1} \approx 3 \times 10^{-14}$~eV, in order to maximize the source contribution.
Henceforth, we will take $m_\phi = 5\e{-15}$~eV as our reference value (see appendix \ref{sec:mphi range} for the impact of different values of $m_\phi$, including possible annual modulation of terrestrial signals).

Then, at the surface of the Earth, we have 
\beq
\phie \approx -\frac{g_n\, N_n^\oplus}{4\pi R_\oplus}e^{-m_\phi R_\oplus}\,.
\label{phi}
\eeq
The above expression is valid as long as we can ignore the $\phi^4$ term in $V(\phi)$ near the Earth's surface which is valid so long as $\lphi<10^{-53}$, as is true for both cases we will consider later.

Let us take $g_n = 5\times 10^{-25}$, consistent with current bounds.
We then have $|\phie|\approx 4 \times 10^{12}$~eV\footnote{Note that $\phi_\oplus$ is accidentally close to the weak scale, and hence $g_s$ ends up near the typical value of Yukawa couplings inferred for the SM neutrinos if the Earth contribution is not negligible.}.
We then take $g_s=5\e{-14}$ so that $-g_s\phie=0.2$ eV.
See Ref.~\cite{Denton:2023iaa} for the details of calculating $\phi$ sourced by a spherical object such as the Earth.
In our scenario, this is the largest fermion coupling of $\phi$, which typically induces a quartic coupling at 1-loop order given by 
\beq
\delta \lphi \sim \frac{g_s^4}{16 \pi^2}\sim 4\e{-56} \left(\frac{g_s}{5\e{-14}}\right)^4\,.
\label{delta-lphi}
\eeq
We mention in passing that for the assumed values of $(g_s, g_n)$, finite 1-loop corrections quadratic in fermion bare masses do not destabilize $m_\phi^2$ (for $m_0 \sim $~eV). 

\section{Neutrino Mass Generation}
In order to address the anomalies ascribed to sterile neutrinos, we need to induce a mixing between active standard model (SM) neutrinos and $\nu_s$.
We have found that this may be achieved in a number of ways, but here we focus on one example with Dirac neutrinos, for concreteness.
Another case with Majorana right-handed neutrinos with more involved phenomenology is sketched in appendix \ref{sec:majorana}.

\subsection{The Dirac Case}
As a specific example of the above scenario, let us take the case of Dirac neutrinos and lepton number conservation.
Lepton number can be stabilized in a number of ways via $U(1)$ or $ Z_n$ type symmetries, see e.g. Ref.~\cite{CentellesChulia:2018bkz}.
We consider the three active flavors $\nu_a$ coupled to three right-handed neutrinos $\nu_R$, as well as an additional neutrino\footnote{While we refer to this new fermion as a neutrino for convenience, it does not carry weak charge.} $\nu_s$ which has both left and right components.
We consider the Yukawa couplings
\beq
-\mathcal L \supset
y_a H \bar \nu_R L_a + 
y_s H \bar \nu_s L_a + {\rm H.C.}\,,
\label{Yukawa}
\eeq
where $y_{a,s}$ are Yukwa coupling constants, $H$ is the SM Higgs doublet, and $L_a$ is the lepton doublet containing $\nu_a$.
The above interactions generate Dirac mass terms $m_\nu = y_a \vev{H}$ and $m_D = y_s \vev{H}$, with $v\equiv \sqrt{2}\vev{H}\approx 246$~GeV the Higgs vacuum expectation value.
We then also have a contribution to the mass of the $\nu_s$ state from \eq{L} $m_s=m_0+g_s\phi$ which changes with the local baryon density.
One can show that the mass eigenvalues are
\begin{align}
m_1 &\simeq m_\nu\frac{m_s}{\sqrt{m_s^2 + m_D^2}}\,,\label{m1}\\
m_{2,3}&\simeq m_\nu\,,\label{m2}
\end{align}
where we note that the heavier of the light states are largely unaffected by $m_s$ and $m_\nu$ is the characteristic size of a $3\times3$ mass matrix.
The heavy state is
\beq
m_4 \simeq \sqrt{m_s^2 + m_D^2 + \frac{m_\nu^2 m_D^2}{m_s^2 + m_D^2}}\,.
\label{m4}
\eeq
The mixing between the active flavors $\nu_a$ and $\nu_4$ is governed by the effective two-flavor angle $\theta_{i4}$, where
\beq
\tan (2 \theta_{i4}) \simeq \frac{2 m_D m_s}{m_s^2-m_D^2-m_\nu^2}\,,
\label{theta-a}
\eeq
for $m_s^2>m_D^2+m_\nu^2$ which is satisfied since $m_s\ge m_0>m_D$.
Generally, $\theta_{i4}$ is required to be $\ord{0.1}$ near the Earth's surface, and implies 
$m_D\gtrsim 0.1$~eV consistent with direct mass searches by KATRIN which finds $m_a\lesssim1$ eV at the Earth's surface \cite{KATRIN:2019yun}.
We set $m_\nu=0.03$ eV and find that $m_D=0.3$ eV and $m_0=1$ eV work well to describe the high significance $>5\sigma$ evidence for sterile neutrino mixing preferring $\Delta m^2_{41}\gtrsim1.25$ eV$^2$ and $\sin^22\theta_{14}\simeq0.34$ from gallium experiments \cite{Barinov:2021asz,Giunti:2022xat}.
We also point out that minor modifications of the parameters can also easily lead to a cosmologically safe explanation of the LSND \cite{LSND:2001aii} and MiniBooNE \cite{MiniBooNE:2020pnu} anomalies while also evading solar neutrino constraints, discussed below.

As mentioned before, the above parameters only describe neutrinos near the surface of the Earth.
However, in environments with much higher matter densities those parameters can be very different.
Note that since the postulated long range force is mediated by a scalar, it does not distinguish between particles and antiparticles.

\section{Cosmology}
A primary motivation for our model is to avoid the constraints on sterile neutrinos that disfavor new light degrees of freedom in the early universe.
However, we will illustrate below how one may also obtain a viable DM candidate in our setup, making our proposal significantly more compelling. 

\subsection{Constraints from Big Bang Nucleosynthesis}

Here, we discuss how our proposal can lead to significant suppression of sterile neutrino production in the early Universe.
Big Bang Nucleosynthesis (BBN) requires a reheat temperature $\trh\gsim 4$~MeV \cite{Hannestad:2004px}.
Hence, we assume that it suffices to show that the model yields consistent phenomenology for $\trh\sim \ord{10\rm~MeV}$, for definiteness and as a minimal proof of principle.
It is straightforward to extend our analysis to higher temperatures.
We take $n_B/s\sim 10^{-10}$, where $n_B$ is the baryon number density and $s\sim g_* T^3$ is the entropy of the Universe; $g_*\sim 10$ is the number of relativistic degrees of freedom at temperatures of interest here.
Typical baryogenesis scenarios are completed by $T\sim \trh$ adopted here and hence we may assume $n_B\sim 10^{-9} T^3$, however the $e^+e^-$ number density $n_e \sim T^3$, until after their annihilation at $T \ll$~MeV, below which $n_e\sim n_B$. 

In order to evade severe constraints from thermalization of $\nu_s$ through neutrino oscillation, we will take the induced mass of $m_s \sim g_s \phi$ to be large compared to its vacuum value $m_0\sim 1$~eV, which in our framework also means a suppressed mixing angle $\theta_{i4}$.
Standard cosmology with $N_{\rm eff}\approx3$ effective neutrinos after BBN consistent with the data is achieved for $m_s\sim$ keV and $\theta_{i4}\sim10^{-3}$ \cite{Hannestad:2015tea,Acero:2022wqg}. We will argue below that our model can easily accommodate the required masses and mixing, in the early Universe.

Let us take $\trh \approx 10$~MeV as a concrete example. We need the right initial scalar value to obtain the above allowed sterile neutrino parameters at 
BBN, corresponding to $T_{\rm BBN}\sim$~MeV, that is: $m_s \gsim$~keV $\Rightarrow$ $\phi_{\rm BBN} \gsim 10^{16}$~eV. 
Allowing for some redshift between $\trh$ and $T_{\rm BBN}$, we require $\phi_i \gsim \text{few} \times 10^{16}$~eV.
As we will explain below, this regime of initial field values $\phi_i$ is required to get the right DM abundance, through the misalignment mechanism (akin to how axion DM would be established in the Universe; see for example Ref.~\cite{Marsh:2015xka}).

As discussed earlier, the 1-loop induced quartic coupling in \eq{delta-lphi} could be naturally given by $\lphi \sim 4\e{-56}$, for our reference choice $g_s = 5\times 10^{-14}$, which would imply the initial dominance of the quartic potential over the mass term.
Note that the mass term starts to dominate at $\phi_*\sim 9\e{13}$~eV.
A scalar dominated by its quartic coupling redshifts like radiation \cite{Turner:1983he} and hence this corresponds to $T_*\sim (\phi_*/\phi_{\rm BBN})T_{\rm BBN} \sim$~keV.
At this temperature, the energy density in the scalar field would be given by $m_\phi^2 \phi_*^2 \sim 0.2$~eV$^4$. This energy density in the oscillating $\phi$ field redshifts like matter, that is $\sim T^3$.
Hence, by the standard epoch of matter-radiation equality at $T\sim$~eV it would be reduced by $\sim$(eV/keV)$^3\sim 10^{-9}$, well below the requisite energy density $\sim$~eV$^4$ to establish the correct DM cosmic energy budget today.

We see that the above ``1-loop" choice for $\lphi$ can be consistent with cosmological constraints on sterile neutrino mass and mixing, but would not explain DM.
We, therefore, consider another regime of parameters that allows us to identify $\phi$ as the dominant DM in the Universe, while also providing acceptable values for $m_s$ and $\theta_{i4}$ before SM neutrino decoupling.
As we will outline below, in this case, the renormalized value of the quartic coupling is small compared to the 1-loop estimate; $\lphi \ll 10^{-56}$.

\subsection{Ultralight Scalar Dark Matter}
\label{sec:uldm}

Let us now consider what parameters can lead to $\phi$ as viable DM, through a initial misalignment $\phi_i$.
We will not explain how the required $\phi_i$ is set, but take it as an input that needs to be realized, through initial cosmological conditions after inflation, or a thermal mechanism \cite{Batell:2021ofv}, in order to get the correct DM abundance. Arguments based on the Milky Way satellite population suggest that the behavior of $\phi$ should transition to matter-like, dominated by its mass term, by the time the Universe has cooled to $T\sim$~keV \cite{Das:2020nwc}.
Henceforth, we will assume this to be the transition temperature $T_{\rm tr}\approx$~keV.
After this point, the scalar energy density would be dominated by $m_\phi^2 \phi^2$ and would redshift like matter.
Thus, by $T_{\rm eq}\approx 1$~eV associated with standard matter-radiation equality era, the energy density in $\phi$ is given by $\sim (T_{\rm eq}/T_{\rm tr})^3\, m_\phi^2 \phi^2/2$.
Hence, at $T_{\rm tr}$, one needs $\phi_{\rm tr}\sim 10^{19}$~eV.
Since prior to $T_{\rm tr}$ the radiation-like quartic term dominates by assumption, we end up with $\phi_i\sim 10^{23}$~eV at $\trh \sim 10$~MeV. We will provide a sketch a modest extension of our scenario that could provide the required misalignment $\phi_i$ in Appendix \ref{sec:misaligned}.

The preceding considerations imply that $\lphi \approx 12m_\phi^2/\phi_{\rm tr}^2\sim 3\times 10^{-66}$ in order for $\phi$ to be viable as the dominant form of DM. 
We will take the above value as a reference for the rest of this work.
Note that for larger $T_{\rm tr}$, one would need a smaller value of $\lphi$ than given above\footnote{In either case considered above, the quartic dominates and hence the value of $\phi$ induced by the nucleon plasma is roughly given by $(g_n n_B/\lphi)^{1/3}\ll \phi_i$, which can, therefore, be neglected compared to the assumed initial value $\phi_i$.}.
One could also consider a slightly different scenario with $\lphi=0$ where $\phi$ is still the dark matter; see appendix \ref{sec:alternative}.

Note that based on the discussion in the last section, the above value of $\phi_i\sim 10^{23}$~eV, together with the reference value $g_s = 5 \times 10^{-14}$, leads to a sufficiently large $m_s\sim 5~\text{GeV} \gg$~keV and small $\theta_{i4}\sim 10^{-10}\ll 10^{-3}$, for consistency with cosmological bounds on $N_{\rm eff}$.
Hence, our ultralight dark matter (ULDM) scenario can result in the correct DM abundance, while satisfying the constraints on sterile neutrino parameters by several orders of magnitude.
The parameters for both scenarios are listed in table \ref{tab:parameters}.

\subsection{Absolute Mass Constraints}
The tightest constraint on the absolute neutrino mass scale comes from combining cosmological data \cite{Planck:2018vyg,DiValentino:2021hoh} and is $\sum_{i=1}^3m_i<0.09$ eV.
Notably, however, this constraint is dominated by data in the $10<z<100$ range \cite{Lorenz:2021alz} when the baryon density was low
and thus $m_s\gtrsim m_0$.
In that limit we find that the mass states are mostly active and also $m_{1,2,3}\to m_\nu$.

\section{Solar Neutrinos}
In the Sun -- particularly in its core -- the properties of $\nu_s$ would be significantly affected by the high densities.
Assuming a mean mass density of $\rho_c^\odot\sim 100$~g/cm$^3$ and a core radius of $R_c^\odot \sim 10 R_\oplus$, the number of nucleons in the core volume is estimated to be $N_c^\odot\sim 7 \times 10^{55}$.
Since $1/R_c^\odot \sim 3 \times 10^{-15}$~eV, the core size can be considered sufficiently large compared to the range of the scalar set by $m_\phi=5\e{-15}$~eV.
The core nucleon number density is then $n_c^\odot \sim 5 \times 10^{11}$~eV$^3$.
Assuming the dominance of the scalar mass term we would then expect $\phi_c^\odot \sim -g_n n_c^\odot/ m_\phi^2 \sim -10^{16}$~eV.
For $\lphi\gsim 3\e{-60}$, this field value implies that the quartic term would be larger than the mass term in the potential of $\phi$.
Hence, one could ignore the mass term and approximate the potential by the quartic term.
For a roughly constant core density, we then find
\beq
\phi_c^\odot \sim -\left(\frac{6 \, g_n \,n_c^\odot }{\lphi}\right)^{1/3} \sim -3\times 10^{14}~\text{eV}\left(\frac{4\e{-56}}{\lphi}\right)^{1/3}\,,
\label{phisun}
\eeq
which yields $m_s^\odot\sim 16$~eV and $\theta^\odot\sim 2\times 10^{-2}$ at the solar core if $\lphi$ dominates.

\begin{figure}
\centering
\includegraphics[width=\columnwidth]{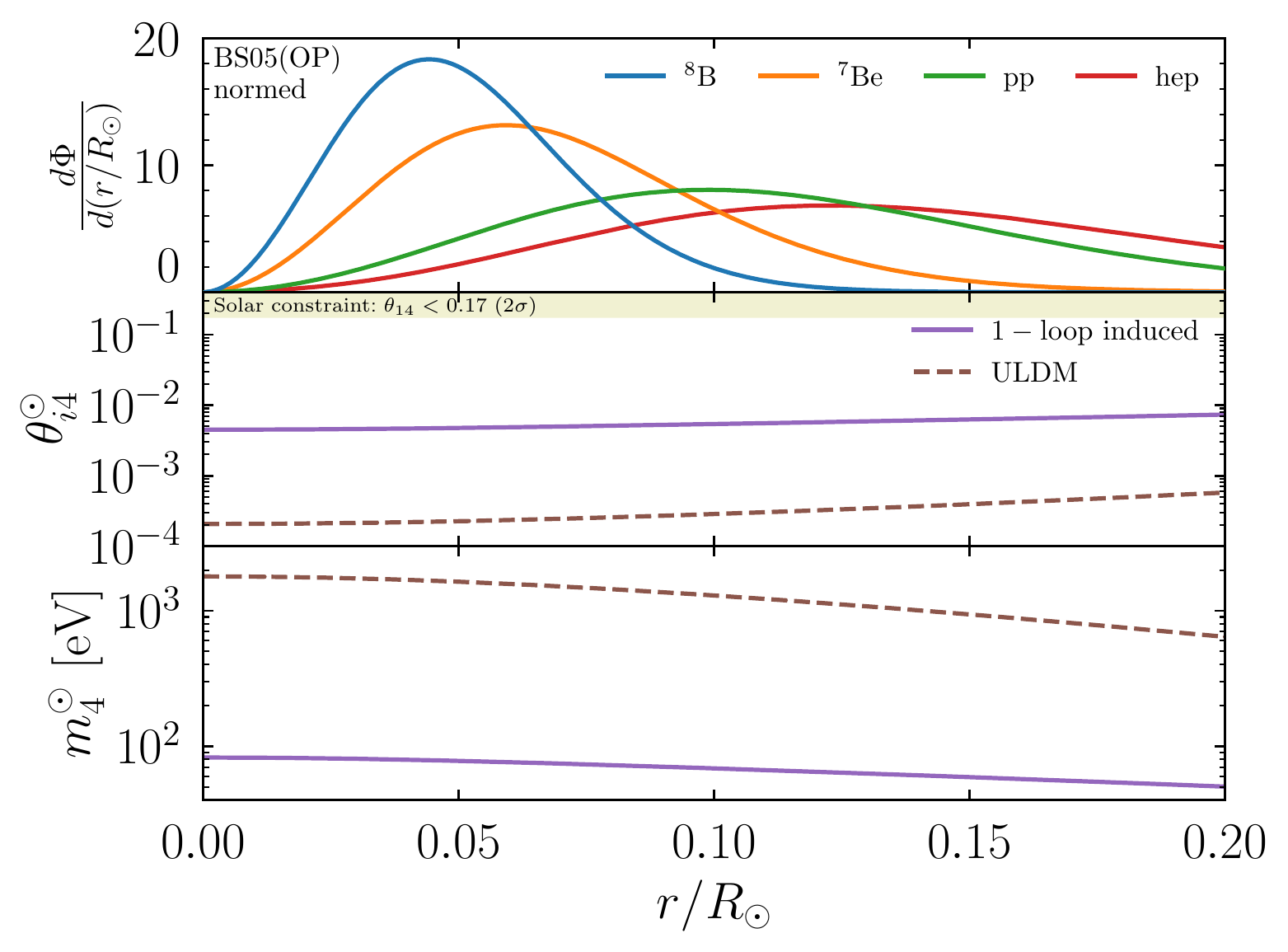}
\caption{\textbf{Top}: The production region in the Sun of the four main processes producing solar neutrinos as a function of radius in the Sun, from Ref.~\cite{Bahcall:2004pz}.
Note that $hep$ neutrinos have not yet been detected.
\textbf{Middle}: The predicted mixing angle in the Sun as a function of radius for the two scenarios described: 1-loop (solid) and ULDM (dashed).
The shaded region shows the mixing parameter values that existing data disfavor \cite{Goldhagen:2021kxe}.
\textbf{Bottom}: The predicted sterile mass in the Sun as a function of radius for the two scenarios described.}
\label{fig:solar}
\end{figure}

Note, however, that the above value for $\phi^\odot_c$ is more than an order of magnitude smaller than the above estimate in the mass-term dominance case.
Hence, we expect that both terms could be important in this regime and the estimate for the sterile neutrino mass and mixing angle would be somewhat larger and smaller, respectively.
This is what we in fact find, as presented in Fig.~\ref{fig:solar} and Table \ref{tab:solar}.
The small induced mixing angles: $\sim5\e{-3}$ in the 1-loop induced $\lphi$ case and $\sim3\e{-4}$ in the ULDM case, allow our scenario to evade solar neutrino constraints quite easily which are independent of mass and disfavor mixing angles $>0.17$ \cite{Goldhagen:2021kxe}.
The much smaller value of $\lphi$ in the ULDM case would lead to $\phi$ mass-term dominance, and even smaller mixing angle.

Using the results in Refs.~\cite{Denton:2023iaa,S5}, we performed a detailed numerical calculation by solving the Klein-Gordon equation with both $m^2\phi^2$ and $\lphi\phi^4$ terms and the Sun's density profile \cite{Bahcall:2004pz} using the initial conditions $\phi(\infty)=0$ and $d\phi/dr|_{r=0}=0$.
We then computed the average mixing angle and sterile mass over the production region of the four most relevant processes: $^8$B, $^7$Be, $pp$, and $hep$.
Our results are shown in Fig.~\ref{fig:solar} and Table \ref{tab:solar}.
Future sensitivity to sterile neutrinos with solar neutrinos \cite{Goldhagen:2021kxe} are not likely to soon reach the levels predicted here, unless $g_s$ is smaller than our fiducial value by $\sim1-2$ orders of magnitude, in which case the $r$ dependence of the sterile signature could conceivably be extracted from the different components of the flux.

Here, we would like to provide an estimate for the value of the coupling $\lambda_*$ which marks the onset of quartic interaction domination.
Assuming a large region with a constant density of charges $n$ coupled to $\phi$ with strength $g$, we find $\lambda_* \sim m_\phi^6/(g^2 n^2)$.
If $\lambda\ll \lambda_*$, we expect the mass term to dominate.
\begin{table}[h]
\centering
\caption{Reference parameters of the two models considered.
The first is where $\phi$ is the DM and the second is where $\lambda$ is at the 1-loop induced value with no additional tuning.}
\begin{tabular}{|l|c|c|}
\hline
& ULDM & 1-loop induced $\lambda_\phi$ \\\hline
$\lambda_\phi$ & $3\e{-66}$ & $4\e{-56}$\\
$g_n$ & $5\e{-25}$ & $5\e{-25}$\\
$g_s$ & $5\e{-14}$ & $5\e{-14}$ \\
$m_\phi$ (eV) & $5\e{-15}$ & $5\e{-15}$\\
$m_0$ (eV) & 1 & 1\\
$m_D$ (eV) & 0.3 & 0.3\\
$m_\nu$ (eV) & 0.03 & 0.03\\\hline
\end{tabular}
\label{tab:parameters}
\end{table}

\begin{table}
\centering
\caption{The effective mixing angle and sterile neutrino mass in the Sun for the four main production regions assuming the benchmark parameters in table \ref{tab:parameters}.
In this mass range the existing constraint is independent of mass and is at $\theta_{14}<0.17$ at $2\sigma$ \cite{Goldhagen:2021kxe}.}
\begin{tabular}{|c|c|c|c|c|}
\hline
& \multicolumn{2}{c|}{ULDM} & \multicolumn{2}{c|}{1-loop induced $\lambda_\phi$}\\\cline{2-5}
& $\theta_{14}$ & $m_4$ (eV) & $\theta_{14}$ & $m_4$ (eV) \\\hline
$^8$B & $2.3\e{-4}$ & $1.6\e3$ & $4.8\e{-3}$ & 77\\
$^7$Be & $2.5\e{-4}$ & $1.5\e3$ & $5.0\e{-3}$ & 74\\
pp & $3.5\e{-4}$ & $1.2\e3$ & $5.6\e{-3}$ & 67\\
hep & $4.2\e{-4}$ & $1.0\e3$ & $5.8\e{-3}$ & 64 \\\hline
\end{tabular}
\label{tab:solar}
\end{table}

One could also investigate the impact of sterile neutrinos on the day-night effect of solar neutrinos \cite{Dooling:1999sg,Palazzo:2011rj,Long:2013ota}.
However, an early analysis of solar neutrino data indicated the role of the Earth on the night time neutrinos applies to a different region of solar neutrino parameter space than is viable given KamLAND data \cite{Giunti:2000wt}.

\section{Supernova Neutrinos}
This scenario will also dramatically modify the behavior of sterile neutrinos inside supernovae.
We find that for both ULDM and 1-loop induced scenarios that the $\lphi$ term contribution to the baryon-induced value of $\phi$ is dominant over the mass term -- although only modestly in the ULDM case -- in the pre-collapse iron core of the supernova.
For a typical radius of $\sim 1000$~km and a density $\sim 10^9$~g/cm$^3$ we get $\phi_c^{\rm SN}\sim-2\e{20}$ eV and thus $m_4\approx m_s\sim8$ MeV and $\theta_{i4}\sim4\e{-8}$ which is safe by many orders of magnitude \cite{Suliga:2019bsq,Suliga:2020vpz}.
If produced, these sterile neutrinos would decay, but at a detectable rate only for mixing angles much larger -- by several orders of magnitude -- than predicted here \cite{Fuller:2008erj,Mastrototaro:2019vug,Syvolap:2023trc}.
For the 1-loop induced case, the $\lphi$ term easily dominates and we find $\phi_c^{\rm SN}\sim-7\e{16}$ eV and thus $m_4\approx m_s\sim 4$~keV and $\theta_{i4}\sim9\e{-5}$ which also easily evade the above supernova bounds.

\section{Atmospheric Neutrinos}
Atmospheric neutrinos at $E\sim1$ TeV are sensitive to sterile neutrino oscillations at $m_4\sim1$ eV through the Earth's core \cite{akhmedov1988neutrino,Krastev:1989ix,Chizhov:1998ug,Chizhov:1999az,Akhmedov:1999va,Petcov:2016iiu}.
IceCube has strong constraints on $\Delta m^2_{41}\sim0.1-1$~eV$^2$ and a weak hint for a sterile neutrino at $\sin^22\theta_{24}\sim0.1$ \cite{IceCube:2020phf}.

We solve the Klein-Gordon equation with both $m_\phi^2$ and $\lphi$ terms in the Earth using the PREM distribution of matter in the Earth \cite{Dziewonski:1981xy}.
We find that, for the parameters of interest, the sterile neutrino mixing angle is smaller and its mass is larger in the Earth's core, but only $\sim10-20\%$ different from the surface values.
In other scenarios this difference could be much larger, and thus detectable (or possibly consistent with IceCube's weak hint); see appendix \ref{sec:phi contribution to masses}.

\section{Conclusions}
In this paper, we have shown that it is possible to have a sterile neutrino that acts as suggested by terrestrial experiments on the Earth's surface, but evades the strong bounds from cosmology.
The required input is extremely minimal: a coupling to an ultralight scalar boson that can be dark matter.
This scenario is compatible with all known cosmological data. 
Moreover, our scenario causes sterile neutrinos to act differently depending on their environment.
For example, the sterile mixing angle is extremely suppressed in the Sun so solar neutrino constraints on sterile neutrinos are no longer competitive.

As the sterile neutrino picture continues to clarify itself in coming years, we point out that this economical scenario predicts a rather different sterile neutrino picture than is usually considered depending on the environment.

\begin{acknowledgments}
We thank J.~Cline for pointing out an error in an earlier version of this manuscript and A.~Suliga for helpful comments. The authors acknowledge support by the United States Department of Energy under Grant Contract No.~DE-SC0012704.
\end{acknowledgments}

\appendix 

\section{Possible Underlying UV Physics}
\label{sec:uvphysics}

In this appendix, we describe one potential picture of the underlying physics that can lead to the various phenomenologically motivated choices we have adopted in our work.
However, we do not attempt to provide a detailed UV complete picture and the following is only meant to provide a sampling of general qualitative possibilities.
We will present several different mechanisms to address various features, but we find that these different mechanisms may be connected to each other as well.

\vspace{0.1in}

In \eq{L}, the $\phi$ couplings to electrons and nucleons could be thought of as originating from higher dimension operators that involve the Higgs field. 
For example, we may have 
\beq
c_\psi\frac{\phi H \bar \psi_L \psi_R}{M_*}\to c_\psi \frac{\vev{H}}{M_*} \phi \bar \psi_L \psi_R\,,
\label{dim5}
\eeq
where $\psi_L$ is an SM $SU(2)$ doublet fermion and $\psi_R$ is a singlet fermion ({\it e.g.}, a right-handed electron or quark); appropriate gauge and Lorentz contraction of indices is assumed in the above {\it schematic} operator. After electroweak symmetry breaking, the coupling to the fermion $\psi$ will be set by $g_\psi \equiv c_\psi \vev{H}/M_*$. For $\psi=e$, this would give our $g_e$ in \eq{L}, but for $\psi$ a light quark, this would yield a coupling to a quark which upon confinement would lead to the nucleon coupling $g_n$.
The coupling parameter $c_\psi$ may be considered as a free parameter, or it may arise out of a deeper theory as discussed below.
The scale $M_*$ is set by the UV physics, and we will come back to it below.

\vspace{0.1in}

Let us now address the choice of structure of the potential of $\phi$.
In \eq{Vphi}, we have ignored the cubic term $\propto \phi^3$. One may simply assume that this is due to a very small strength for this interaction.
Alternatively, we could more reasonably assume that there is a $Z_2$ symmetry under which both $\phi$ and the left-handed chirality of the sterile neutrino $\nu_{sL}$ are odd which automatically forbids the $\phi^3$ term but allows the $\phi\bar\nu_{sL}\nu_{sR}$ term. 
 However, the bare mass term proportional to $m_0$ is now disallowed.

The resolution to forbidden $m_0\bar\nu_{sL}\nu_{sR}$ term could be provided in an extra dimensional theory. Here, we do not speculate on the size and detailed properties of the extra dimensions. However, any successful formulation of string theory generally requires compact extra spatial dimensions which provides possible motivation for the scenario we will outline here. In such a theory, various fields can be localized along the extra dimensions and on various ``branes," {\it i.e.}, sub-spaces in the larger manifold \cite{Arkani-Hamed:1999ylh,Grossman:1999ra,Gherghetta:2000qt}.

In light of the above considerations, let us assume that there is another scalar $\chi$ that has the same $Z_2$ charge as $\phi$ and $\nu_{sL}$.
We assume that the sterile neutrino $\nu_{sL}$ and $\phi$ are localized on the same brane while the scalar $\chi$ lives on another brane. We also posit that $\chi$ develops an expectation value and locally break the $Z_2$. 
Hence, one may have a mass term $m_0$ generated by the expectation value of $\chi$, as long as $\nu_{sL}$ and $\nu_{sR}$ have some overlap at the location of $\chi$. 
Yet, by locality along the extra dimension, $\phi$ and $\chi$ may not have any appreciable interactions, assuming complete localization of the scalars on separate branes. Thus, no local $\vev{\chi} \phi^3$ term may be generated in the 4-dimensional effective theory which allows us to continue ignoring the $\phi^3$ term.

We will not provide a detailed scheme for the localization of various fields. However, given that there can be several dimensions where our field content can reside, one could in general generate the hierarchies among couplings assumed in our work, through localization along those dimensions.

Note though that the coupling of $\phi$ to electrons and nucleons violates the assumed $Z_2$. This could be addressed in a 2-Higgs doublet model, where the additional Higgs $H'$ field has the same $Z_2$ charge as $\phi$. This avoids possible requirement of a gravitational instanton to violate the $Z_2$ charge \cite{Kallosh:1995hi,Calmet:2019frv}, which could lead to sever exponential suppression of $c_\psi$ in \eq{dim5}. Nonetheless, the extreme smallness of the requisite $g_{e,n}$ couplings argues for a quantum gravity generated operator, corresponding to $M_*$ being identified as the scale of gravity, {\it i.e} the Planck or string scale. With the replacement $H\to H'$ in \eq{dim5}, and assuming $\vev{H'}\lsim 100$~GeV, one could achieve effective $g_{e,n}\sim 10^{-17}\,c_\psi$. This suggests $c_\psi\lsim 10^{-7}$, which may be due to small overlap of the fields along the extra dimensions.

\section{Initial Misalignment Mechanism}
\label{sec:misaligned}
Here, we would like to outline a moderate expansion of our model that would yield the requisite $\phi_i\sim 10^{23}$~eV at $T\sim 10$~MeV, for $\phi$ to be ULDM, as described in section \ref{sec:uldm}. This can be achieved if we assume a reheat temperature $\trh\gsim 100$~MeV, and that muons also couple to $\phi$, with a strength $g_\mu \sim 10^{-19}$, which is allowed by current bounds \cite{Davoudiasl:2018ltz}. A rough estimate yields 
\beq
\phi_i \sim \left(\frac{g_\mu T^3}{\lambda_\phi}\right)^{1/3} \sim 5\times 10^{23}~\text{eV}\,,
\label{phii}
\eeq
from interactions with muons.
As $T$ drops below the mass of muons, $\sim 106$~MeV, $\phi_i$ is no longer supported by the thermal ensemble of $\mu^\pm$ and starts to oscillate with roughly the value needed to establish ULDM by $T\sim 10$~MeV. The above suffices to show that one could in principle have a dynamical mechanism to set $\phi_i$ near values assumed in our work.
See also Ref.~\cite{Batell:2021ofv} for discussions of scalar misalignment using similar effects.

\section{The Majorana Case}
\label{sec:majorana}
One could consider an alternative case with Majorana mass terms for the right-handed neutrinos.
This could be realized in several ways, one of which we outline here.
Then the required terms in the Lagrangian are
\begin{equation}
-\mathcal L\supset y_a H\overline{\nu_R} L_a+
\xi \phi\, \overline{\nu_s}\nu_R + 
\frac12m_R\overline{\nu_R^c}\nu_R+{\rm H.C.}\,,
\end{equation}
which is effectively equivalent to the minimal extended type I seesaw \cite{Barry:2011wb} except that a mixing between the sterile state and $\nu_R$ is sourced by the $\phi$ field.
Here, we have assumed that a possible $\phi \overline{\nu_s^c} \nu_s$ term, akin to one included in \eq{L} is negligibly small.

Assuming that $m_R$ is the largest mass scale in the setup, we can integrate out $\nu_R$, as in the conventional seesaw models.
We then get the following interactions in the low energy effective theory
\beq
g_a \phi\, \overline{\nu_s}\nu_{La} + \frac{\xi^2 \phi^2}{m_R} \overline{\nu_s^c}{\nu_s} + {\rm H.C.},
\label{EFT}
\eeq
where $g_a \equiv \xi\, y_a \vev{H}$. Note that $y_a \vev{H}$ is the Dirac mass term associated with active flavor $a$, in a conventional seesaw mechanism.

The above terms in the low energy theory can lead to an alternative scenario where both the mass of the sterile state and its mixing with the active states vanish away from sources of $\phi$, {\it i.e.}~empty space.
In that case, one would retrieve the standard active masses and mixing angles.
Note that the sterile-active mixing angle $\theta_{i4} \propto 1/\phi$ and thus when $\phi$ is large, as in the early Universe or in dense astrophysical environments, the mass of the sterile state becomes large, while $\theta_{i4}$ gets small, and hence typical constraints from cosmological or Solar data can be addressed.
The phenomenology can be rather more involved than what is discussed in the main text since in the limit where $\xi\phi$ transitions from small to large relative to $m_R$ or when $\xi\phi$ transitions from large to small relative to $y_a\langle H\rangle$, the active neutrino mass, the hierarchy will change.
We leave the exploration of this scenario to future work.

\section{Sun as Source and Annual Modulations}
\label{sec:annual modulation}
There are other interesting features of this model with slightly different parameters (not necessarily leading to $\phi$ as DM).
For example, if $m_\phi$ is smaller by several orders of magnitude than $5\e{-15}$ eV from table \ref{tab:parameters}, then the dominant contribution at the Earth would be from the Sun.
Thus, there would be no spatial variation of the sterile parameters in different terrestrial experiments, but solar neutrino experiments would see nothing unexpected as the sterile mixing angle in the Sun would be dramatically suppressed.
There would, however, be an annual modulation signature in the $g_s\phi$ contribution to any terrestrial sterile neutrino data as the distance $r$ from the Earth to the Sun varies by $3.4\%$ with a peak in July and a minimum January (for a similar effect in different contexts, see Ref.~\cite{Davoudiasl:2011sz,Picoreti:2015ika}).
Assuming the $m_\phi$ term dominates over the $\lphi$ term in \eq{Vphi}, we find that
\begin{equation}
\frac{\Delta\phi}\phi=-(m_\phi r+1)\frac{\Delta r}r\,.
\end{equation}
This could manifest itself as Lorentz Invariance Violation (LIV).
Several searches for LIV with neutrinos have been performed \cite{MINOS:2010kat,SNO:2018mge,Arguelles:2021kjg,DayaBay:2018fsh}, but, to our knowledge, none have looked for this specific effect.

\section{ULDM \texorpdfstring{\boldmath$\phi$}{phi} Contribution to Neutrino Masses}
\label{sec:phi contribution to masses}
If $\phi$ constitutes cosmic DM, it has a background value that furnishes finite active and sterile neutrino masses.
In the Galactic vicinity of the Solar System, the DM energy density is given by $\rho_{\phi; \rm MW\odot} \sim 0.3$~GeV cm$^{-3}$ \cite{ParticleDataGroup:2022pth}.
Assuming the reference mass $m_\phi \sim 5\times 10^{-15}$~eV, away from astronomical bodies we have $\phi_{\rm MW\odot} \sim 4 \times 10^{11}$~eV -- which oscillates with a frequency given by $m_\phi$ -- corresponding to $g_s \phi_{\rm MW\odot} \sim 2 \times 10^{-2}$~eV.
By contrast, today's cosmic DM energy density is given by $\rho_0 \sim 4 \times 10^{-11}$~eV \cite{ParticleDataGroup:2022pth}, and hence the cosmological value $\phi_0 \sim 2 \times 10^9$~eV corresponding to $g_s \phi \sim 10^{-4}$~eV is obtained for our reference parameters, which is small enough.

Using the results in the main text, if $m_0=0$, the sum of neutrino masses $\sum_i m_i \sim m_D\sim 0.3$~eV, well above current cosmological bounds \cite{DiValentino:2021hoh}.
This is due to the smallness of the sterile mass induced by DM and hence we need a bare mass $m_0\gsim 1$~eV, as assumed in the main text, to comply with constraint on the sum of neutrino masses in empty space.

\section{Resonant Sterile Neutrino Production in the Early Universe}

In the early universe, the sterile neutrino parameters can be different if the background dark matter field is set by $\phi$, as described in the main text.
Such a classical field would oscillate and, importantly, pass through a resonance which could produce a larger number of sterile neutrinos than naively expected.

We consider some benchmark numbers to show that this sterile neutrino production does not affect $N_{\rm eff}$.
Following from Eqs.~\eqref{m1}-\eqref{m4}, the resonance ($\theta_{i4}=45^\circ$) happens at $m_s=m_D$ (we can safely take $m_\nu=0$).
From Ref.~\cite{Acero:2022wqg}, we have that for $m_s=0.3$ eV one needs $\theta_{i4}\lesssim0.06$ to ensure that $\Delta N_{\rm eff}\lsim 0.1$.
We find that $\theta_{i4}>0.06$ when\footnote{There is a tiny region near $m_s=0$ where $\theta_{i4}$ is small. We ignore this region.} $|m_s|<6.6$ eV. We ignore the bare mass $m_0\sim1$ eV as it simply offsets the above region by a small amount.
Thus we are interested in values of $|\phi|<\phi_c$ near $\phi_c \sim 10^{14}$ eV and we focus on the ULDM case with $\lphi\sim 3\times 10^{-66}$.

We now determine the amount of time in the early universe spent when $|\phi|<\phi_c$.
Since the $\lambda_\phi\phi^4/{4!}$ term generally dominates in the $T\sim$MeV regime, the full solution for such an oscillating field is non-trivial, but can be easily solved in the regime of interest.
We note that the $m_\phi^2\phi^2/2$ term dominates the energy density when $\phi\lsim 10^{19}$ eV which is certainly true for the region of interest below $\phi_c$.
The total energy density in the field at $T=1$ MeV is $10^{21}$ eV$^4$ from the $\lambda_\phi$ term\footnote{Note that the amplitude of $\phi$ evolves as $\phi\propto T$ in the $\lambda_\phi$ dominated regime.}.
Near $\phi=0$ the solution is a simple harmonic oscillator which has a solution $\phi(t)=A\sin(m_\phi t)$ for some $A$.
Since the total energy must be conserved, it must be equal to the kinetic energy density $\frac12\dot\phi^2=\frac12A^2m_\phi^2$, thus $A=10^{25}$ eV.
Note that the effective amplitude in this mass dominated regime is several orders of magnitude larger than the actual amplitude.
The interpretation of this is that the behavior of the field with both $m_\phi$ and $\lambda_\phi$ terms in the regime that is dominated by the mass term is equivalent to a field with no $\lambda_\phi$ term and a different amplitude.
Now we compute the time spent in the large mixing angle region as
\begin{equation}
t_{\rm res}=\frac2{m_\phi}\frac{\phi_c}A\sim 4\e{3}\text{ eV}^{-1}\sim 3\e{-12}\text{ s}\,,
\end{equation}
where we have safely used the small angle approximation.

Next, we compute the oscillation time at $T=1$ MeV.
Oscillations to sterile neutrinos will occur according to
\begin{equation}
P_{as}=\sin^22\theta_{i4}\sin^2\left(\frac{\Delta m^2_{41}t}{4E}\right)\,.
\end{equation}
We have that $\Delta m^2_{41}<m_4^2\approx 2m_D^2 \approx 0.2$ eV$^2$ in the interesting region.
As $\phi,m_s\to0$, $\Delta m^2_{41}$ decreases more which will only increase the oscillation time.
Then, we have $m^2_{41}/(4E)\lsim 5\e{-8}$ eV.
Plugging in the numbers from above, we find that while the mixing angle may be large $\sim0.1-\pi/4$, there will not ever be enough time for oscillations to occur as $\sin^2(\Delta m^2_{41}t/4E)\lsim 4\e{-8}$.

Next, we must ensure $|\phi|\lsim \phi_c$ is not satisfied for a large fraction of the relevant Hubble time, leading up to neutrino decoupling at $T\sim1$ MeV, to avoid $\Delta N_{\rm eff}$ values in conflict with existing constraints.
We find that the period of $\phi$ increases from $\gsim 10^{10}$ eV$^{-1}$ at $T=10$ MeV to $\gsim 10^{12}$ eV$^{-1}$ at 1 MeV.
Thus the field only spends $<10^{-8}$ of the time in the resonant production state ensuring that the sterile neutrinos are not over produced.

\section{An Alternative DM Scenario}
\label{sec:alternative}
Here, we outline an alternative scenario that avoids some of the subtleties of the reference parameter space, for which $\phi$ oscillates before neutrino decoupling at $T\lsim 1$~MeV, as discussed in the previous section of the appendix.
We will assume that $\lphi=0$ and that $V(\phi)$ only consists of $m_\phi^2 \phi^2/2$.
To hold the field at its initial value in the early universe, we will set its mass to be smaller than the Hubble rate $H \sim 10^{-15}$~eV at $T\sim 1$~MeV.
For example, we can have $m_\phi \sim 10^{-17}$~eV, which still ensures that the value of background $\phi$ at terrestrial experiments is dominated by nucleons from the Earth.
As discussed before, with $m_s\gsim $~keV and $\theta_{i4}\lsim 10^{-3}$ we can avoid cosmological constraints on sterile neutrinos.
Hence we may choose the initial value $\phi_i\gsim 4\times 10^{16}$~eV, for $g_s \sim 5\times 10^{-14}$.

With the above choices, the values of sterile neutrino mass and mixing stay at the allowed levels until $\phi$ starts to oscillate at $T\sim 0.1$~MeV, corresponding to the $H \sim m_\phi$, after which the energy density $\rho_\phi=m_\phi^2 \phi_i^2/2$ redshifts like matter.
For $\rho_\phi$ to have the standard $\sim$~eV$^4$ DM value at $T\sim 1$~eV, we then need to have $\phi_i\sim 3\times 10^{24}$~eV (incidentally, not far from interesting ultraviolet scales associated with grand unification or string theory).
Note that for the above $\phi_i$ value, $m_s \gg$~keV and $\theta_{i4}\ll 10^{-3}$ until after SM neutrinos have decoupled and sterile neutrino thermalization via oscillation has turned off.

\section{Allowed Ranges of the Mass of the Mediator}
\label{sec:mphi range}
We have chosen the mediator to have a mass of $m_\phi=5\e{-15}$ eV, but other masses would also work.
First, increasing the mass is viable, but it will decrease the strength of the field near the Earth's surface and in the Sun.
While decreasing $\phi$ near the Earth's surface will not affect the oscillation physics much, in the Sun it will start to decrease the mass, and thus increase the mixing angle.
This could be accommodated by increasing $g_s$ ($g_n$ is already near the limit from fifth force probes and the limit on $g_e$ is tighter than that on $g_n$ so it will not significantly contribute).
Larger $g_s$ makes the 1-loop induced size of the $\lambda_\phi$ term larger potentially requiring more cancellation in the dark matter scenario.
Without increasing $g_s$ we find that increasing $m_\phi$ from its fiducial value essentially guarantees that it dominates the potential in the Sun and decreases the value of the field, thus increasing the mixing angle.
We find that we need $m_s>1.75$ eV to ensure consistency with the Solar constraint, this in turn requires $|\phi^\oplus_c|>1.5\e{13}$ eV and thus $m_\phi<10^{-13}$ eV.
This limit can be relaxed by increasing $g_s$ subject to the above caveats.

Second, decreasing the mediator mass is also viable.
At $m_\phi\sim10^{-18}$ eV, $1/m_\phi$ is approximately the distance from the Sun to the Earth and the Sun will start to contribute to the potential at the Earth, although this does not change much, except for the small annual modulation signature, see \ref{sec:annual modulation}.
For $m_\phi\sim10^{-20}$ eV to $m_\phi\sim10^{-21}$ eV, if $\phi$ is the dark matter, additional benefits and constraints come into play.
Measurements of the Lyman-$\alpha$ forest disfavor sufficiently light dark matter in this range \cite{Rogers:2020ltq}, while small scale structure data may actually prefer dark matter in this mass range \cite{Bullock:2017xww}.
In the early universe, small values of $m_\phi$ will require some care with regards to the values of $\lphi$ to ensure that it redshifts like matter if it is the dark matter.

\bibliography{main}

\end{document}